\documentclass[prd,onecolumn]{revtex4}
\usepackage{dcolumn}
\usepackage{multirow}
\usepackage{graphicx}
\usepackage{amssymb}
\usepackage{bm}
\usepackage{hyperref}
\usepackage{epstopdf}
\usepackage{color}
\usepackage{mathrsfs}
\usepackage{amsmath,amssymb,amsthm}
\begin{document}
\title{A new pressure-parametrization unified dark fluid model}
\author{Deng Wang$^{1}$}
\email{Cstar@mail.nankai.edu.cn}
\author{Yang-Jie Yan$^{2}$}
\email{yanyj926@gmail.com}
\author{Xin-He Meng$^{2}$}
\email{xhm@nankai.edu.cn}

\affiliation{
$^1${Theoretical Physics Division, Chern Institute of Mathematics, Nankai University, Tianjin 300071, China}\\
$^2${Department of Physics, Nankai University, Tianjin 300071, China}\\}
\begin{abstract}
We propose a new pressure-parametrization model to explain the accelerated expansion of the late-time Universe by considering the dark contents (dark matter and dark energy) as a unified dark fluid. To realize this model more physically, we reconstruct it with the quintessence and phantom scalar fields, respectively.     We use the recent cosmological data to constrain this model, distinguish it from the standard cosmological model and find that the value of the Hubble constant $H_0=68.34^{+0.53}_{-0.92}$ supports the global measurement by the Planck satellite at the $1\sigma$ confidence level.
\end{abstract}
\maketitle
\section{Introduction}
With the ever-growing data, modern cosmological observations such as Type Ia supernovae (SNe Ia), baryonic acoustic oscillations (BAO), cosmic microwave background (CMB) anisotropies, observational Hubble parameter (H(z)), etc., have confirmed the fact that our Universe is undergoing a phase of accelerated expansion \cite{1,2,3,4}. The best way to explain such an evolution is the simple addition of the cosmological constant (CC) or $\Lambda$ term \cite{5} in the framework of general relativity (GR). However, this otherwise formally and observationally consistent model meets two unsolved puzzles, i.e., the so-called fine-tuning and coincidence problems \cite{6}. The former indicates that the enormous disagreement between the energy scale introduced by the CC and the predictions of standard model of particle physics for the vacuum energy density, while the latter that implies the $\Lambda$-cold-dark-matter ($\Lambda$CDM) model can not explain why the Universe is only recently accelerating. As a consequence, to alleviate or even solve these problems, a flood of alternative cosmological models based on different physical origins are proposed and studied by cosmologists. In the literature, there are mainly two classes of models, i.e., the well-known dark energy (DE) models, which introduce a new matter component in the cosmic pie, and the theories of modified gravities (MOG), which deviate from the standard GR. Here we only list a part of them as follows: phantom \cite{7}, quintessence \cite{8,9,10,11,12,13,14,15}, bulk viscosity \cite{16,17,18,19,20,21}, decaying vacuum \cite{22,23,24}, Chaplygin gas \cite{25}, f(R) gravity \cite{26,27,28,29,30,31}, Einstein-Aether gravity \cite{32,33}, braneworld gravity \cite{34,35,36} and so on.

As is well known, the parametrization method can be regarded as a powerful and useful tool to characterize the properties of the DE component in the late-time Universe. This method has been widely applied into confronting cosmological observations in recent years. Generally speaking, to investigate the properties and features of DE component, the mainstream of the DE parametrization is the equation of state (EoS) parametrization. One can express this class of parametrization in a polynomial form $\omega(z)=\sum\limits_{n=0}\omega_nx_n(z)$, where the expansions could be provided as follows: (i) $x_n(z)=z^n$; (ii) $x_n(z)=(\frac{z}{1+z})^n$; (iii ) $x_n(z)=[\ln(1+z)]^n$; (iv) $x_n(z)=[\frac{z(1+z)}{1+z^2}]^n$, etc. The parametrization (i) was firstly propoesd by Huterer et al. \cite{37} and Weller et al. \cite{38} investigate the case of $n\leqslant1$. Although this parametrization behaves very well at low redshifts, it exhibits a problematic behavior at high redshifts such as failing to explain the age estimations of high-$z$ (z denotes the redshift) objects, since it predicts substantially small ages at $z\geqslant3$. The so-called Chevalier-Polarski-Linder (CPL) parametrization \cite{39} (ii) with the case of $n\leqslant1$ are dedicated to solving the above-mentioned problematic behavior at high redshifts, which fits well for a great deal of theoretically conceivable scalar field potential, and provide an excellent explanation for small deviations from the phantom barrier ($\omega=-1$). The parametrization (iii) with the case of $n\leqslant1$ was introduced by G. Efstathiou \cite{40},  which is intended to adjust some quintessence-like scenarios at $z\lesssim4$. The parametrization (iv) was proposed by Barboza et al. \cite{41}, which is aimed at extending the range of applicability of the DE EoS, and avoid the uncertainties and singularities in the aforementioned three scenarios. Recently, several new parameterizations have been proposed, such as applying the Pad\'{e} method into the DE EoS \cite{42}.

About ten years ago, A. A. Sen proposed a parametric scenario for the effective pressure of the DE, namely $P_{\Lambda}=-P_0+P_1(1-a)+\cdot\cdot\cdot$ so as to study the small deviations from the $\Lambda$CDM model \cite{43}, where $a$ is the scale factor. Following this theoretical line, we have developed two parametric models for the effective pressure of the unified dark fluid (UDF), i.e., $P(z)=P_a+P_bz$ and $P(z)=P_c+\frac{P_d}{1+z}$ \cite{44}, and found these two models fit current observations well. The two scenarios both indicate the DE EoS $\omega<-1$, which corresponds to a phantom-like case at the present epoch. Actually, one can easily find that this class of parametrization is equivalent to the above-mentioned class through some simple derivations. In this situation, to continue studying the underlying DE model deviating from the $\Lambda$CDM model, we propose a new unified pressure-parametrization dark fluid model, investigate its evolutional behavior, place constraints on it using the recent cosmological data and exploring its corresponding value of the Hubble constant $H_0$.

The rest of this study is organized in the following manner. In the next section, we propose the new model. In Sec. 3, we reconstruct it with quintessence and phantom scalar fields, respectively. In Sec. 4, we constrain this model by using the recent cosmological observations. In Sec. 5, we distinguish this model from the $\Lambda$CDM model. In Sec. 6, we investigate the $H_0$ value of our model. The discussions and conclusions are presented in the final section (we take units $8\pi G=c=1$).

\section{The model}
The Friedmann equations, the conservation equation of the stress-energy tensor, and the EoS compose a close dynamical system to characterize the background evolution of the Universe. As mentioned above, the EoS is equivalent to the relation between the effective pressure and the redshift. For an spatially flat   Friedmann-Robertson-Walker (FRW) spacetime, on the one hand, one can obtain the $P(z)-z$ relation by inserting the EoS into the equation of energy conservation. On the other hand, the EoS can be recovered by inserting the $P(z)-z$ relation into the equation of energy conservation, which is expressed as
\begin{equation}
\dot{\rho}+3H(\rho+P)=0, \label{1}
\end{equation}
Notice that the $\Lambda$CDM scenario is not strictly supported by current cosmological observations since there exists a small deviation. However, the physical mechanism about the small deviation is still not clear. Different from the scenarios including an extra component and the evolution of the EoS, we would like to regard the dark content in the late-time Universe as a UDF, and investigate the parametrization scenario for effective pressure of the UDF which actually circumvents the issue of possible physical mechanism. Utilizing this approach, we can study the possible deviations from the constant case of $P(z)-z$ relation without a specific presupposition.

The new pressure-parametrization scenario as a UDF can be written as
\begin{equation}
P(z)=P_a+P_b(z+\frac{z}{1+z}), \label{2}
\end{equation}
where $P_a$ and $P_b$ are two different free parameters. Replacing Eq. \ref{2} in Eq. \ref{1}, we have
\begin{equation}
\rho(a)=-P_a+\frac{3}{4}P_b(a-2a^{-1})+Ca^{-3}, \label{3}
\end{equation}
where $C=\rho_0+P_a+\frac{3P_b}{4}$ is an integration constant and $\rho_0$ is the present-day energy density. Note that the subscript `` 0 '' represents the present values of the corresponding physical quantities and we have used the usual relation $a=\frac{a_0}{1+z}=\frac{1}{1+z}$ here. It is interesting to see that in Eq. (\ref{3}), the constant term $P_a$ and the last term $Ca^{-3}$ can be regarded as the CC term and the dust matter term in the $\Lambda$CDM model, respectively. The second term $\frac{3}{4}P_b(a-2a^{-1})$ can be interpreted as the small deviations from the CC.

Subsequently, dividing on both sides of Eq. (\ref{3}) by $\rho_0$ and using the first Friedmann equation $H^2=\frac{\rho}{3}$, we obtain the dimensionless Hubble parameter as
\begin{equation}
E(a)=\{\alpha+\beta(2a^{-1}-a)+(1-\alpha-\beta)a^{-3}\}^{\frac{1}{2}}, \label{4}
\end{equation}
where $E(a)=H(a)/H_0$, and $\alpha=-P_a^*$ and $\beta=-\frac{3}{4}P_b^*$ ($P_a^*=P_a/\rho_0$, $P_b^*=P_b/\rho_0$) are two free parameters to be constrained by observations in the following section. To exhibit the scalar field reconstructions of our model better, we derive the EoS of our model as follows
\begin{equation}
\omega_{de}=-1+\frac{\beta(2a^{-1}+a)}{3[\alpha+\beta(2a^{-1}-a)]}. \label{a1}
\end{equation}

\section{The reconstructions}
In principle, the deviations from the $\Lambda$CDM in our model can be realized by different physical scenarios. The scalar field reconstructions are main methods to explain the accelerated expansion of the Universe at the present epoch. To perform a detailed analysis in the context of reconstructions, one must investigate the evolutional behaviors of the scalar fields and potential over cosmic history, and compare them with those of energy density and pressure of our pressure-parametrization UDF model at low redshifts. The specific realization of the scalar field reconstructions can be expressed as
\begin{equation}
P_{eff}=P_{scalar}, \label{5}
\end{equation}
\begin{equation}
\rho_{eff}=\rho_{scalar}+\rho_{m}, \label{6}
\end{equation}
where $\rho_m$ is the matter density in the late-time Universe. In this work, we would like to reconstruct our model with quintessence and phantom scalar fields, respectively.

The quintessence ($\omega_{de}>-1$) corresponds to a canonical scalar field $\phi$ with a potential $V(\phi)$ that only interact with the standard gravity, and its action is written as
\begin{equation}
S=\int d^4x \sqrt{-g}[\frac{R}{2}-\frac{1}{2}g^{\mu\nu}\partial_{\mu}\phi\partial_{\nu}\phi-V_1(\phi)]+S_m, \label{7}
\end{equation}
where $R$ is the Ricci scalar and $S_m$ is the action of matter. By the variation of the lagrangian of quintessence with respect to (w.r.t.) $\phi$, one can obtain
\begin{equation}
\ddot{\phi}+3H\dot{\phi}+V'_1(\phi)=0, \label{8}
\end{equation}
where the dot and the prime denote the derivatives w.r.t. the cosmic time and the scalar field $\phi$, respectively. In a spatially flat FRW Universe, the energy density and pressure of the quintessence scalar field can be expressed as
\begin{equation}
\rho_{de}=\frac{1}{2}\dot{\phi}^2+V_1(\phi), \label{9}
\end{equation}
\begin{equation}
P_{de}=\frac{1}{2}\dot{\phi}^2-V_1(\phi). \label{10}
\end{equation}
It follows that the EoS is
\begin{equation}
\omega_{de}=\frac{\frac{1}{2}\dot{\phi}^2-V_1(\phi)}{\frac{1}{2}\dot{\phi}^2+V_1(\phi)}. \label{11}
\end{equation}
The phantom corresponds to a scalar field which is minimally coupled to the standard gravity and matter sources, and its action is
\begin{equation}
S=\int d^4x \sqrt{-g}[\frac{R}{2}+\frac{1}{2}g^{\mu\nu}\partial_{\mu}\phi\partial_{\nu}\phi-V_2(\phi)]+S_m. \label{12}
\end{equation}
The dynamically evolutional equation of the phantom field can be written as
\begin{equation}
\ddot{\phi}+3H\dot{\phi}-V'_2(\phi)=0. \label{13}
\end{equation}
Furthermore, the energy density, pressure and EoS of the phantom field can be, respectively, shown as
\begin{equation}
\rho_{de}=-\frac{1}{2}\dot{\phi}^2+V_2(\phi), \label{14}
\end{equation}
\begin{equation}
P_{de}=-\frac{1}{2}\dot{\phi}^2-V_2(\phi), \label{15}
\end{equation}
\begin{equation}
\omega_{de}=\frac{-\frac{1}{2}\dot{\phi}^2-V_2(\phi)}{-\frac{1}{2}\dot{\phi}^2+V_2(\phi)}. \label{16}
\end{equation}

\subsection{The Quintessence Case}
\begin{figure}
\centering
\includegraphics[scale=0.5]{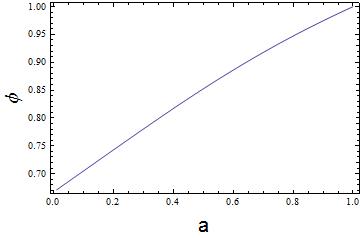}
\includegraphics[scale=0.5]{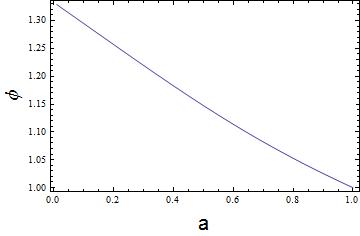}
\includegraphics[scale=0.5]{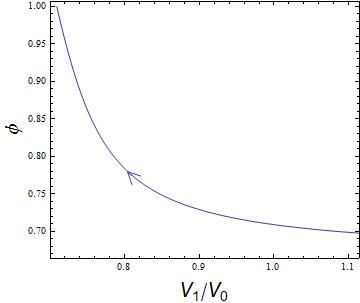}
\includegraphics[scale=0.5]{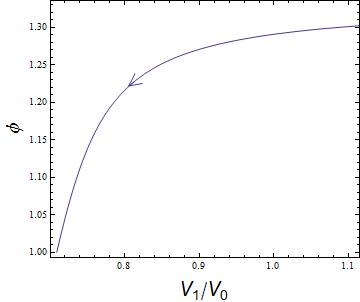}
\caption{The upper left and lower left panels corresponding to a plus sign in Eq. (\ref{21}) represent the relation between the quintessence field $\phi$ and the scale factor $a$, and the relation between the quintessence field $\phi$ and the potential $V_1/V_0$. The upper right and lower right panels corresponding to a minus sign in Eq. (\ref{21}) represent the relation between the quintessence field $\phi$ and the scale factor $a$, and the relation between the quintessence field $\phi$ and the potential $V_1/V_0$. The arrows indicate the evolutional direction of the potential. We have assumed $\alpha=0.7$ and $\beta=0.02$ numerically.}\label{f1}
\end{figure}

Assuming that the cosmic pie consists of two ingredients (quintessence and matter) and using Eqs. (\ref{2}-\ref{3}) and Eqs. (\ref{14}-\ref{15}), one can easily obtain
\begin{equation}
\frac{1}{2}\dot{\phi}^2+V_1(\phi)=-P_a+\frac{3}{4}P_b(a-2a^{-1}),  \label{17}
\end{equation}
\begin{equation}
\frac{1}{2}\dot{\phi}^2-V_1(\phi)=P_a+P_b(a^{-1}-a),  \label{18}
\end{equation}
Simplifying the above two equations with the redefined parameters $\alpha$ and $\beta$, we have
\begin{equation}
\frac{1}{2}\dot{\phi}^2=\frac{1}{6}\rho_0\beta(a+2a^{-1}),  \label{19}
\end{equation}
\begin{equation}
V_1(\phi)=\rho_0[\alpha+\frac{1}{6}\beta(10a^{-1}-7a)].  \label{20}
\end{equation}
In order to solve Eqs. (\ref{19}-\ref{20}) conveniently, we use the initial condition $\phi_{a=1}=M_{pl}$, where $M_{pl}$ is the reduced Planck mass. Then the first Friedmann equations can be rewritten as $H^2=\frac{1}{3M_{pl}}\rho$. Considering the dark energy dominated Universe at the present stage with the density ratio parameter $\Omega_{de}\sim0.7$, we can define the present value of the potential as $V_0=\rho_0=3M_{pl}^2H_0^2$. Subsequently, by solving Eqs. (\ref{19}-\ref{20}), one can obtain
\begin{equation}
\frac{d\phi}{da}=\pm M_{pl}[\frac{\beta(a^{-1}+2a^{-3})}{\alpha+\beta(2a^{-1}-a)+(1-\alpha-\beta)a^{-3}}]^{\frac{1}{2}},  \label{21}
\end{equation}
\begin{equation}
V_1(\phi)=V_0[\alpha+\frac{1}{6}\beta(10a^{-1}-7a)],  \label{22}
\end{equation}
where the symbol `` $\pm$ '' denotes two different solutions. We would like to study the evolutional behaviors of the quintessence fields numerically by considering $\alpha=0.7$ and $\beta=0.02$. The results are presented in Fig. \ref{f1}. One can find that $\phi$ increases with the increasing scale factor $a$ in the upper left panel of Fig. \ref{f1}, and that the potential decreases monotonically with the increasing $\phi$ and it will reach the minimal value at the recent future $z\approx-5.3\times10^{-10}$ (see the lower left panel of Fig. \ref{f1}), which implies the Universe will eventually tend to be a de Sitter one. In the upper right panel of Fig. \ref{f1}, one can find $\phi$ decreases with the increasing $a$, and in the lower right panel of Fig. \ref{f1}, the potential decreases monotonically with the decreasing $\phi$ and also reaches its minimal value at the recent future. Furthermore, using Eq. (\ref{4}), we can obtain the density ratio parameter $\Omega_{\phi}$ of the quintessence field as

\begin{figure}
\centering
\includegraphics[scale=0.5]{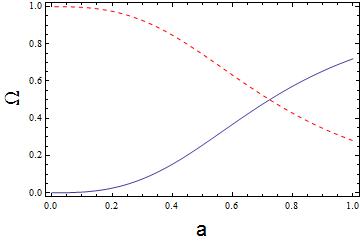}
\caption{The evolutional behaviors of the density ratio parameters for the quintessence field (solid line) and the matter (dashed line). }\label{f2}
\end{figure}
\begin{figure}
\centering
\includegraphics[scale=0.5]{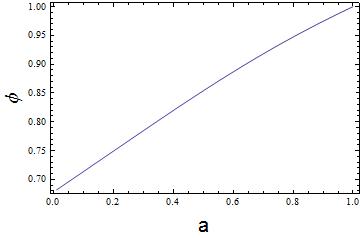}
\includegraphics[scale=0.5]{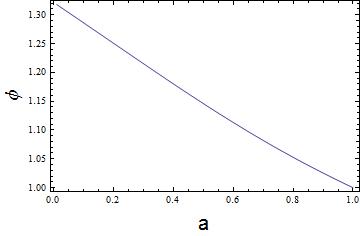}
\includegraphics[scale=0.5]{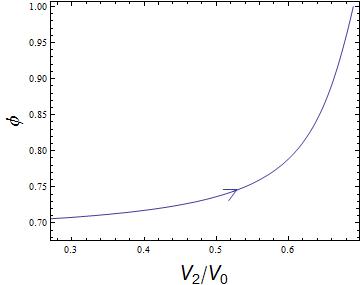}
\includegraphics[scale=0.5]{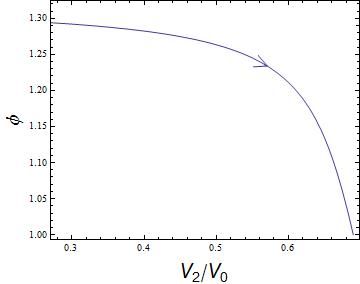}
\caption{The upper left and lower left panels corresponding to a plus sign in Eq. (\ref{28}) represent the relation between the phantom field $\phi$ and the scale factor $a$, and the relation between the phantom field $\phi$ and the potential $V_2/V_0$. The upper right and lower right panels corresponding to a minus sign in Eq. (\ref{28}) represent the relation between the phantom field $\phi$ and the scale factor $a$, and the relation between the phantom field $\phi$ and the potential $V_2/V_0$. The arrows indicate the evolutional direction of the potential. We have assumed $\alpha=0.7$ and $\beta=-0.02$ numerically.}\label{f3}
\end{figure}
\begin{figure}
\centering
\includegraphics[scale=0.5]{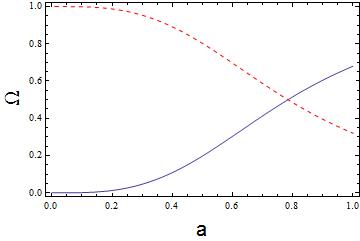}
\caption{The evolutional behaviors of the density ratio parameters for the phantom field (solid line) and the matter (dashed line). }\label{f4}
\end{figure}

The evolutional tendency of the density ratio parameter $\Omega_{\phi}$ of the quintessence field is exhibited in Fig. \ref{f2}, which is cosmologically dominant until low redshifts and is well consistent with the $\Lambda$CDM model at low redshifts.

\subsection{The Phantom Case}
Utilizing Eqs. (\ref{2}-\ref{3}) and Eqs. (\ref{14}-\ref{15}) and assuming the cosmic pie consists of two ingredients (phantom and matter), one can obtain
\begin{equation}
-\frac{1}{2}\dot{\phi}^2+V_2(\phi)=-P_a+\frac{3}{4}P_b(a-2a^{-1}),  \label{24}
\end{equation}
\begin{equation}
-\frac{1}{2}\dot{\phi}^2-V_2(\phi)=P_a+P_b(a^{-1}-a),  \label{25}
\end{equation}
Similarly, simplifying the above two equations with the redefined parameters $\alpha$ and $\beta$, we have
\begin{equation}
\frac{1}{2}\dot{\phi}^2=-\frac{1}{6}\rho_0\beta(a+2a^{-1}),  \label{26}
\end{equation}
\begin{equation}
V_2(\phi)=\rho_0[\alpha+\frac{1}{6}\beta(10a^{-1}-7a)].  \label{27}
\end{equation}
Utilizing Eq. (\ref{26}), we find the free parameter $\beta<0$. Combining Eqs. (\ref{26}-\ref{27}) with model parameters $\alpha$ and $\beta$, we obtain the following two equations
\begin{equation}
\frac{d\phi}{da}=\pm M_{pl}[-\frac{\beta(a^{-1}+2a^{-3})}{\alpha+\beta(2a^{-1}-a)+(1-\alpha-\beta)a^{-3}}]^{\frac{1}{2}},  \label{28}
\end{equation}
\begin{equation}
V_2(\phi)=V_0[\alpha+\frac{1}{6}\beta(10a^{-1}-7a)].  \label{29}
\end{equation}
According to Eq. (\ref{23}), here we explore the evolutional behaviors of the quintessence fields numerically by considering $\alpha=0.7$ and $\beta=-0.02$. In the upper left panel of Fig. \ref{f3}, one can find that the phantom field $\phi$ increase monotonically with $a$, and that  in the lower left panel of Fig. \ref{f3}, the potential increases monotonically with the increasing $\phi$ and it will reach its maximal value at the recent future. From the upper right panel of Fig. \ref{f3}, one can find $\phi$ decreases with the increasing $a$, and in the lower right panel of Fig. \ref{f3}, the potential increases monotonically with the decreasing $\phi$ and also reaches its maximal value at the recent future. This implies that the phantom field will come to rest at the maximum of the potential and the Universe tends to be a de Sitter one. By investigating the evolutional tendency of the density ratio parameter, we find the phantom field is cosmologically dominant at low redshifts and is well consistent with the predictions of the $\Lambda$CDM model.

\section{The constraints}
In this section, we adopt the Markov Chain Monte Carlo (MCMC) method to constrain the pressure-parametrization UDF model by using SNe Ia, BAO and CMB observations. We use the publicly available package CosmoMC \cite{45,46} to estimate the cosmological parameters that describe the data best. In order to obtain converging results, we stop sampling by checking the worst e-values (the variance of chain means/mean of variances) and $R-1$ is of order 0.01 in the MCMC calculations.

The observations of SNe Ia provide an powerful tool to probe the expansion history of the Universe. As is well known, the absolute magnitudes of all the SNe Ia are considered to be the same, since all the SNe Ia almost explode at the same mass ($M\approx-19.3\pm0.3$). In light of this reason, SNe Ia can theoretically be used as the standard candles to constrain different cosmological models. In this analysis, we use the `` Joint Light-curve Analysis '' (JLA) containing 740 SNe Ia data points, which covers the redshift range $z \in [0.01, 1.3]$ \cite{47}. JLA data can be divided in to four classes: 118 low-$z$ SNe in the range $z \in [0, 0.1]$ from \cite{48,49,50,51,52,53}; 374 SNe in the range $z \in [0.3, 0.4]$ from the Sloan Digital Sky Survey (SDSS) SNe search \cite{54};  239 SNe in the range $z \in [0.1, 1.1]$ from the Supernova Legacy Survey (SNLS) project \cite{55}; 9 high-$z$ SNe in the range $z \in [0.8, 1.3]$ from the Hubble Space Telescope (HST) \cite{56}.

The baryon acoustic oscillations in the primordial plasma have striking effects on the anisotropies of the CMB and the large-scale structure of matter. We also use four different BAO measurements: the 6dFGS at effective redshift $z_{eff}=0.106$ \cite{57}, the SDSS-MGS at $z_{eff}=0.15$ \cite{58}, the BOSS-LOWZ at $z_{eff}=0.32$ and the CMASS-DR11 at $z_{eff}=0.57$ \cite{59}.

\begin{figure}
\centering
\includegraphics[scale=0.5]{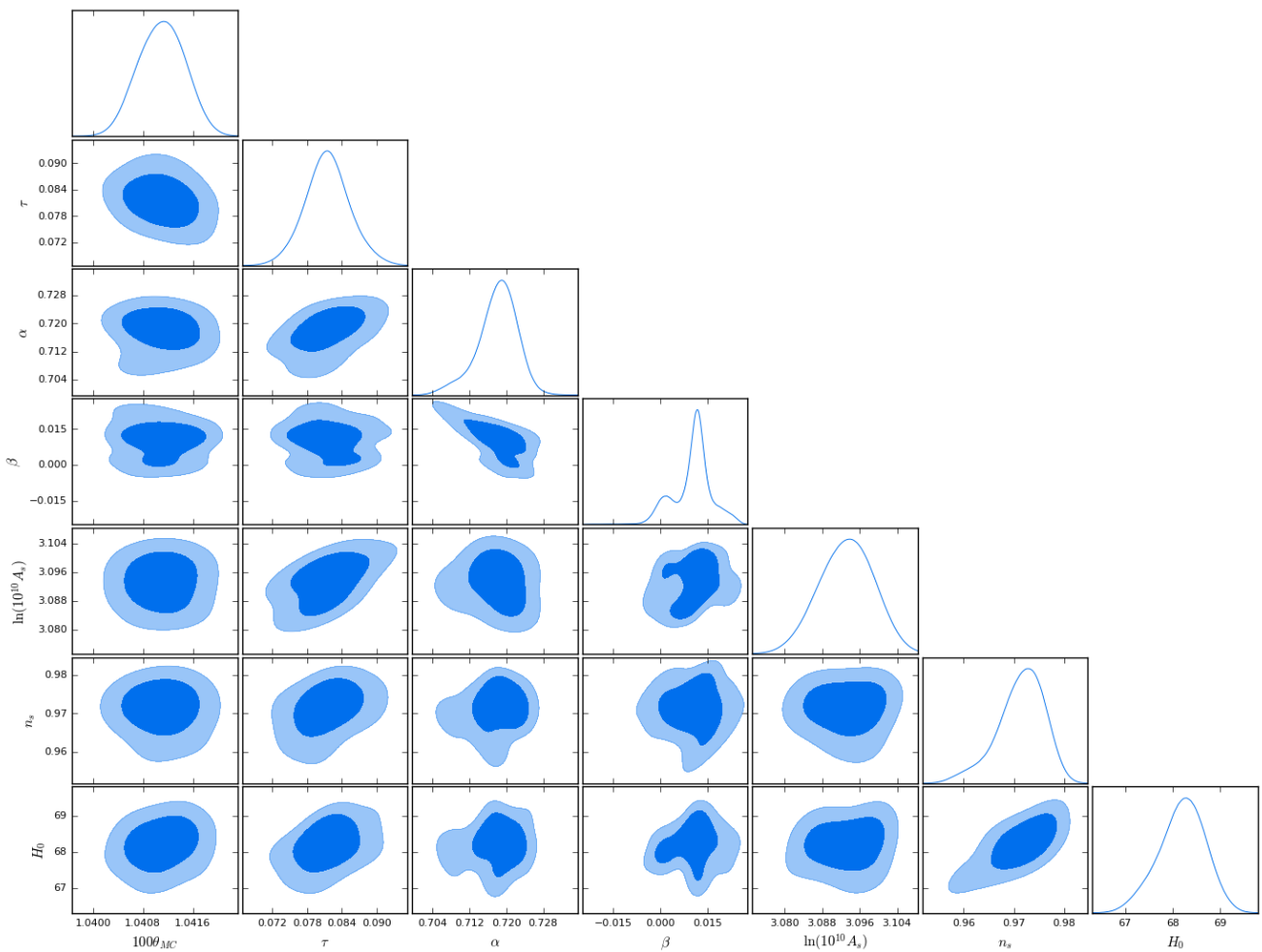}
\caption{The $1$-dimensional marginalized probability distribution on the individual parameter and $2$-dimensional contours of our UDF model by using SNe + BAO +CMB observations.}\label{f5}
\end{figure}

\begin{table}[h!]
\begin{tabular}{ll}
\hline
\hline
      parameter                     &prior    \\
\hline
$100\theta_{MC}$            &$[0.5, 10]$                      \\
$\tau$                    &$[0.01, 0.8]$                           \\
$\alpha$                    &$[0.5, 0.9]$                         \\
$\beta$                    &$[-0.1, 0.1]$                                \\
$\mathrm{ln}[10^{10}A_s]$        &$[2.0, 4.0]$         \\
$n_s$                   &$[0.8, 1.2]$                          \\
$H_0$                   &$[40, 100]$                           \\
\hline
\hline
\end{tabular}
\caption{The priors of different parameters used in the posterior analysis.}
\label{t1}
\end{table}
\begin{table}[h!]
\begin{tabular}{lc}
\hline
\hline
      parameter                     &best-fit value with $1\sigma$ error    \\
\hline
$100\theta_{MC}$            &$1.0411^{+0.0006}_{-0.0007}$                      \\
$\tau$                    &$0.082^{+0.005}_{-0.007}$                            \\
$\alpha$                    &$0.718^{+0.004}_{-0.005}$                         \\
$\beta$                    &$0.011^{+0.006}_{-0.013}$                                 \\
$\mathrm{ln}[10^{10}A_s]$        &$3.093^{+0.007}_{-0.008}$          \\
$n_s$                   &$0.973^{+0.005}_{-0.007}$                           \\
$H_0$                   &$68.34^{+0.53}_{-0.92}$                           \\
\hline
\hline
\end{tabular}
\caption{The priors of different parameters used in the posterior analysis.}
\label{t2}
\end{table}
\begin{equation}
\Omega_{\phi}=\frac{\alpha+\beta(2a^{-1}-a)}{\alpha+\beta(2a^{-1}-a)+(1-\alpha-\beta)a^{-3}}.  \label{23}
\end{equation}
\begin{figure}
\centering
\includegraphics[scale=0.38]{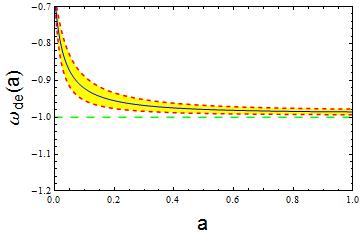}
\includegraphics[scale=0.38]{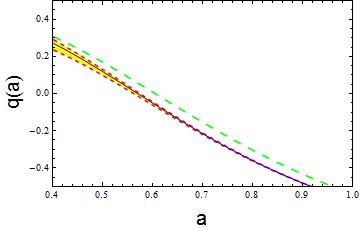}
\includegraphics[scale=0.38]{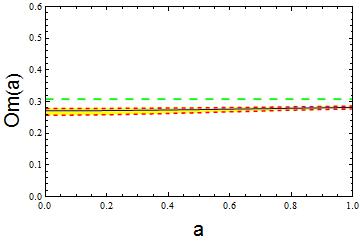}
\caption{From left to right, we exhibit the relations between the scale factor $a$ and the DE EoS $\omega(a)$, deceleration parameter $q(a)$ and $Om(a)$ diagnostic, respectively. The solid (blue) lines and long-dashed (green) lines correspond to the pressure-parametrization UDF model and $\Lambda$CDM model, respectively. The shaded (yellow) regions and short-dashed (red) lines represent the $1\sigma$ regions and corresponding boundaries.}\label{f6}
\end{figure}

As is well known, the afterglow of the big bang---the CMB not only provides some useful information on the very early Universe, but also can give
information on the expansion history of the Universe through the positions of the CMB acoustic peaks. In our analysis, we also use the CMB data from Planck-2015 \cite{60}, i.e., a joint observations of lensing + lowTEB + low$l$ + TT temperature fluctuations angular power spectrum (see the online Package \cite{61}).

To perform the so-called $\chi^2$ statistics, we choose the total likelihood function $\mathcal{L}\propto e^{-\chi^2/2}$ as the product of the separate likelihoods of SNe Ia, BAO and CMB data. Then the total $\chi^2$ can be expressed as
\begin{equation}
\chi^2=\chi^2_{SNe}+\chi^2_{BAO}+\chi^2_{CMB},  \label{30}
\end{equation}
with the corresponding 6-dimensional parameter space
\begin{equation}
\{100\theta_{MC}, \quad \tau, \quad  \alpha, \quad  \beta, \quad  \mathrm{ln}[10^{10}A_s], \quad  n_s\}.  \label{31}
\end{equation}
To implement the constraints, we list the priors on the parameter space in Tab. \ref{t1}. In Fig. \ref{f5}, we exhibit the $1$-dimensional marginalized probability distribution on the individual parameter and $2$-dimensional contours of our UDF model by using SNe + BAO +CMB observations. Meanwhile, the best-fit value and $1\sigma$ error of the model parameters are presented in Tab. \ref{t2}. One can find that the value of $\alpha$ is about $\alpha\sim0.7$, which approaches the contribution of the CC in the $\Lambda$CDM model, and the value of $\beta$ is constrained to be very small. This implies our UDF model deviates slightly from the $\Lambda$CDM model in a point of view of data.

\section{The diagnostics}
In this section, we are full of interest to distinguish the new UDF model from the $\Lambda$CDM model by using the DE EoS $\omega(a)$, deceleration parameter $q(a)$ and so-called $Om(a)$ diagnostic \cite{62}. The expression of the DE EoS $\omega(a)$ is shown in Eq. (\ref{a1}) and the deceleration parameter $q(a)$ is written as
\begin{equation}
q(a)=-\frac{a}{E(a)}\frac{dE(a)}{da}-1,  \label{32}
\end{equation}
where $E(a)$ of the model is shown in Eq. (\ref{4}).

The $Om(a)$ diagnostic is an useful method to distinguish various kinds of cosmological models from the $\Lambda$CDM model and one from the other, which can be defined as
\begin{equation}
Om(a)=\frac{E^2(a)-1}{a^{-3}-1}.  \label{32}
\end{equation}
It is easy to see that the $Om(a)$ diagnostic involves only the first derivative of the scale factor through the Hubble parameter and is easier to reconstruct from the observed data. For a spatially flat $\Lambda$CDM model, it is simply
\begin{equation}
Om(a)=\Omega_{m0},  \label{33}
\end{equation}
where $\Omega_{m0}$ denotes the present-day value of the matter density ratio parameter.

Using $\Omega_{m0}=0.308$ \cite{60} for the flat $\Lambda$CDM model and error propagations of the model parameters $(\alpha, \beta)$, we exhibit the above three diagnostics of the UDF model in Fig. \ref{f6}. One can find that the UDF model deviates slightly from the $\Lambda$CDM model at low redshifts and it will approach gradually the $\Lambda$CDM model in the future. This indicates that the small deviation from the $\Lambda$CDM model has been realized by our new UDF model, and it may provide a new possibility to explain the puzzles, which the standard cosmology meets.

\section{The $H_0$ tension}
Recently, the improved local measurement $H_0=73.24\pm1.74$ km s$^{-1}$ Mpc$^{-1}$ from Riess et al. 2016 (hereafter R16) \cite{63} exhibits a stronger tension with the Planck 2016 release $H_0=66.93\pm0.62$ km s$^{-1}$ Mpc$^{-1}$ (hereafter P15) \cite{64} at the $3.4\sigma$ level. In this analysis, we try to explain it by utilizing the new UDF model deviating slightly from the $\Lambda$CDM model. We obtain $H_0=68.34^{+0.53}_{-0.92}$, which is consistent with the P15's result at the $1\sigma$ level (see Fig. \ref{f5} and Tab. \ref{t2}).

\section{Discussions and Conclusions}
Parametrization approaches for the effective pressure of the dark contents of the Universe are aimed at exploring various kinds of possible small deviations from the $\Lambda$CDM model, in order to alleviate or even solve the problems which exist in the $\Lambda$CDM model. Comparing with the parametrization approaches for the DE EoS, it is interesting to investigate the deviations from the constant $P(z)-z$ relation without any prejudice to a presupposition.

In this work, we propose a new pressure-parametrization model by regarding the dark contents of the Universe as a unified dark fluid. To realize the new UDF model more physically, we reconstruct it with the quintessence and phantom scalar fields, and find that the Universe eventually tends to be a de Sitter type. Subsequently, we constrain the UDF model using the SNe Ia, BAO and CMB observations, and find that the value of $\alpha$ is about $\alpha\sim0.7$, which approaches the contribution of the CC in the $\Lambda$CDM model, and the value of $\beta$ is constrained to be very small. This implies that our UDF model deviates slightly from the $\Lambda$CDM model in a point of view of data. Furthermore, utilizing the constrained model parameters $(\alpha, \beta)$, we distinguish the UDF model from the $\Lambda$CDM model by the DE EoS $\omega(a)$, deceleration parameter $q(a)$ and so-called $Om(a)$ diagnostic. We find that the UDF model deviates slightly from the $\Lambda$CDM model at low redshifts and it will approach gradually the $\Lambda$CDM model in the future. This indicates that the small deviation from the $\Lambda$CDM model has been realized by our new UDF model, and it may provide a new possibility to explain the problems and puzzles, which the $\Lambda$CDM meets.

We also try to explain the current $H_0$ tension by utilizing the UDF model deviating slightly from the $\Lambda$CDM model. The value of the Hubble constant $H_0=68.34^{+0.53}_{-0.92}$ is consistent with the P15's result at the $1\sigma$ confidence level (see Fig. \ref{f5} and Tab. \ref{t2}).

In a follow-up study, we explore some possible parametrization models for the effective pressure.
For example, (i) $P(z)=P_a+P_b\ln(1+z)$: this model leads to a small logarithmic deviation from the the $\Lambda$CDM model; (ii) $P(z)=P_a+P_b\frac{z(1+z)}{1+z^2}$ or $P(z)=P_a\sin(\frac{1}{1+z})$: interesting periodic deviations occur in these two models; (iii) $P(z)=P_a+P_b(1+z)^2+P_c(1+z)^4$: it is interesting that the the $\Lambda$CDM model will be recovered in this model, containing four different evolution epochs of the Universe. The detailed analysis about these pressure-parametrization UDF models are in preparation.

\section{acknowledgements}
This study is supported in part by the National Science Foundation of China. We thank Profs. S. D. Odintsov and Bharat Ratra for useful communications on cosmology and theories of gravity. Deng Wang warmly thanks Prof. Jing-Ling Chen for beneficial discussions.

\end{document}